\documentclass[twocolumn,english,pra,showpacs]{revtex4}
\usepackage[T1]{fontenc}
\usepackage[latin9]{inputenc}
\usepackage{verbatim}
\usepackage{amsmath}
\usepackage{graphicx}
\usepackage{amssymb}
\usepackage{esint}

\makeatother
\usepackage{babel}

\begin{document}

\title{Dynamics of the inhomogeneous Dicke model}

\author{Oleksandr Tsyplyatyev}

\author{Daniel Loss}

\affiliation{Department of Physics, University of Basel, Klingelbergstrasse 82,
CH-4056 Basel, Switzerland }

\date{\today}

\begin{abstract}
We study the time dynamics of a single boson coupled to a bath
of two-level systems (spins 1/2)  with different excitation energies, described by an inhomogeneous
Dicke model. Analyzing the time-dependent Schr\"odinger equation exactly we find that at resonance the boson 
decays in time to an oscillatory state with a finite amplitude characterized by a single Rabi frequency if the inhomogeneity is below a certain threshold.
In the limit of small inhomogeneity, the decay is suppressed and exhibits  a complex  (mainly Gaussian-like) behavior, whereas
the decay  is complete and of  exponential form in the opposite limit. For intermediate inhomogeneity, the boson decay is partial and
governed by a combination of exponential and power laws.
\end{abstract}


\maketitle
\section{Introduction}

Coherent interaction between light and matter \cite{Gardiner}
continues to receive strong interest 
due to significant experimental progress in various areas of physics.
Prime examples are the achievement of Bose-Einstein
condensation of cold-atom gases in electromagnetic traps \cite{cold_atoms_BEC} which
made possible the coherent coupling of $10^{5}$ atoms to a single photon
of an optical resonator \cite{kimble, esslinger}. The time dynamics of quantum optical systems
has received particular attention
\cite{Rempe, Bloch} due to fast optical probing techniques,
especially in the context of quantum metrology based on cavity-QED systems containing 
atomic ensembles \cite{Wineland,Kasevich}.
Advances in
solid-state technology enabled the fabrication of optical microcavities in semiconductors
where electron-hole excitations 
in quantum wells
are strongly coupled to a photon eigenmode of the cavity \cite{kasprzak,ep3}.
Strong coupling of  a transmission-line resonator
to a Cooper-pair box \cite{wallraff} 
as well as  coupling of a cavity to a  single semiconductor quantum dot have been demonstrated\cite{Berezovsky,Badolato1}.
Several schemes for quantum computing based on light-matter interaction have been proposed \cite{Cirac,AwschalomQED,Lukin,Burkard,Trifa:2008}.

The theoretical understanding of all these coupled light-matter systems is based on a model
introduced long ago by Dicke \cite{Dicke},  which describes $N$  two-level
systems ('spin bath') with excitation energies $\epsilon_{j}$ coupled to a single boson
mode  $\omega$ of the quantized light-field, see Eq. ~(\ref{eq:Dicke_model}) below. 
For the special case of identical atoms ($\epsilon_{j}=\epsilon$) and constant couplings constant $g_{j}$ between boson and spin bath this model has been diagonalized \cite{TavisCummings}, and the time dynamics 
obtained exactly \cite{CummingsDorri}.
For inhomogeneous $g_{j}$ (but still constant $\epsilon_{j}$) 
the boson  was shown to oscillate
with a single Rabi frequency $\Omega=\sqrt{N\left\langle g^{2}\right\rangle }$,
where $\sqrt{\left\langle g^{2}\right\rangle} $
is an effective spin coupling. Also  perturbative \cite{Chumakov} and numerical \cite{Solano} approaches
to the time dynamics were considered.

In this paper we solve the quantum time-dynamics of a 
single boson mode coupled to a bath of {\it non-identical} spins 1/2 characterized by  
inhomogeneous energy ('Zeeman') splittings $\epsilon_{j}$ with bandwidth $\Delta$. In condensed matter systems such energy inhomogeneities are generally expected, a typical example being
the exciton-polariton system where such inhomogeneities arise from the unavoidable disorder in a semiconductor \cite{ExcitonPolaritonsEstimate}. In quantum optical systems atomic levels are usually quite perfect ($\epsilon_j\equiv \epsilon$); however, for example, in  cold-atom QED systems such inhomogeneities can play a role as a trap induces spatial variation of the magnetic field \cite{ColdGasesEstimate}.

Analyzing the time-dependent Schr\"odinger equation exactly we find that the bosonic occupation number decays only partially if the inhomogeneity $\Delta$ is below a threshold given by a single Rabi frequency $\Omega$. Below the threshold the boson decays
to an oscillatory
state determined by $\Omega$ and a reduced amplitude which
decreases with increasing ratio $\Delta/\Omega$.
The time decay is exponential for large spin-bath inhomogeneity  $\Delta\gg\Omega$, is complex (mainly Gaussian-like) in the opposite limit $\Delta\ll\Omega$
and is a combination
of exponential and power law behavior in  the intermediate regime $\Delta\simeq\Omega$.
These results are valid if the  boson energy  is tuned 
in resonance with the average spin excitation energy $\left\langle \epsilon\right\rangle -\omega=0$.
With increasing detuning $\left|\left\langle \epsilon\right\rangle -\omega\right|\gg\textrm{max}\left\{\Omega,\Delta\right\}$
the time dynamics of the boson becomes suppressed.

The paper is organized as follows. In Section \ref{II} we analyze the time-dependent Schr\"odinger equation and derive the exact solution in the Laplace domain. In Section \ref{III} we consider rectangular and Gaussian distribution functions of $\epsilon_j$ in resonance with the boson mode, $\langle \epsilon \rangle=\omega$, to obtain the time evolution of the wave functions. Section \ref{IV} contains the analysis and discussion of a finite detuning, $\langle \epsilon \rangle\neq\omega$. In the appendices we give details of the calculations in Sections \ref{III} and \ref{IV}.

\section{The Inhomogeneous Dicke model\label{II}}
The Hamiltonian for the Dicke model governing the dynamics of a single boson mode coupled to $N$ two-level systems is given by
\begin{equation}
H=\omega b^{\dagger}b+\sum_{j=1}^{N}\epsilon_{j}S_{j}^{z}+\sum_{j=1}^{N}g_{j}\left(S_{j}^{+}b+
S_{j}^{-}b^{\dagger}\right)\label{eq:Dicke_model}\, ,\end{equation}
where
 $S_{j}^{\alpha}$
 are spin-1/2 operators, $S_{j}^{\pm}=S_{j}^{x}\pm iS_{j}^{y}$,
and $b\left(b^{\dagger}\right)$  the standard Bose annihilation
(creation) operator \cite{applicability}.
The total number of excitations,
$L=n+\sum_{j}S_{j}^{z}$, is conserved
in the Dicke model,
where ${n}=b^{\dagger}b$ is the bosonic occupation number.
The eigenvalues $c$ of $L$ are the so-called cooperation numbers, 
given by $c=\left\langle L\right\rangle$, where $\left\langle... \right\rangle$ denotes the
expectation value.

In the following we assume that the spin bath can be prepared in its ground state with all spins down, e.g. either dynamically or by thermal cooling \cite{superradiance}.
Also, the  mode $\omega$ is assumed to be  empty or occupied by one boson only.
The non-equilibrium dynamics of a single boson excitation can then be initiated
by a short radiation pulse from an external source. The dissipation
of the boson mode, e.g. through leakage of photons through the mirrors
that define an optical cavity, can be used to detect the dynamics
if the cavity escape time exceeds the internal
time scales of the system dynamics. In a multi-shot experiment  \cite{Rempe,Bloch} the probability
of detecting a leaking photon at a given time is proportional to the
boson expectation value. 
Next, we note that if initially the system has only one excitation,
either in the spin or in the boson subsystem, the subsequent time evolution is restricted to
this subspace and described by the general state
\begin{equation}
\left|\Psi\left(t\right)\right\rangle=\alpha\left(t\right)\left|\Downarrow,1\right\rangle +\sum_{j=1}^{N}\beta_{j}\left(t\right)\left|\Downarrow\uparrow_{j},0\right\rangle \end{equation}
with $c=-N/2+1$, and where $\alpha\left(t\right)$ and  $\beta_{j}\left(t\right)$
are normalized amplitudes, $\left|\alpha\left(t\right)\right|^{2}+\sum_{j}\left|\beta_{j}\right|^{2}=1$,
 of finding  either a state with one boson and no spin excitations present
or a state with no boson and the $j^{\textrm{th}}$-spin excited (flipped) \cite{centralspin}. 

The time evolution within this subspace is determined by the interaction term
in Eq. (\ref{eq:Dicke_model}) that transfers back and forth the excitations between the spin bath and the
boson. Inserting $\left|\Psi\left(t\right)\right\rangle$
 into
the time-dependent Schr\"odinger equation we obtain 
\begin{eqnarray}
-i\frac{d\alpha\left(t\right)}{dt} & = & -\sum_{j}\frac{\left(\epsilon_{j}-\omega\right)}{2}\alpha\left(t\right)+\sum_{j}g_{j}\beta_{j}\left(t\right),\label{eq:eqs_of_motion}\\
-i\frac{d\beta_{k}\left(t\right)}{dt} & = & \sum_{j}\left(\epsilon_{j}-\omega\right)\left(\delta_{jk}-\frac{1}{2}\right)\beta_{k}\left(t\right)+g_{k}\alpha\left(t\right).\nonumber \end{eqnarray}
In above derivation we have subtracted the integral of motion $\omega L$
from the Hamiltonian Eq. (\ref{eq:Dicke_model}) as
it leads only to an  overall phase of $ \left|\Psi\right\rangle$
with no observable effect. The initial conditions $\alpha\left(0\right)=1$, $\beta_{j}\left(0\right)=0$ assumed in the following
correspond to a singly occupied boson mode. The physical observable of interest is
the time-dependent expectation value of the
boson occupation number, which can be expressed in terms of the amplitude $\alpha$ as
$\left\langle n\left(t\right)\right\rangle =\left\langle\Psi\left(t\right) \right| n\left|\Psi\left(t\right)\right\rangle=
|\alpha\left(t\right)|^{2}$.

The set of equations, Eq.  (\ref{eq:eqs_of_motion}), is equivalent to the one obtained in the Weisskopf-Wigner theory in the study of bosonic systems \cite{YamamotoImamoglu}  in contrast of spins 1/2 considered here.  We solve 
Eq. (\ref{eq:eqs_of_motion}) by making use of the Laplace transform, $\alpha\left(s\right)=\int_{0}^{\infty}dt\alpha\left(t\right)e^{-st}$, $\Re{s}>0$.
In the Laplace domain we obtain then a system of linear algebraic equations.
By solving them we find 
\begin{equation}
\alpha\left(s\right)=\frac{i}{is+\frac{N\left\langle \omega-\epsilon_{j}\right\rangle }{2}-\left\langle \frac{g_{j}^{2}N}{i s+\left\langle \omega-\epsilon_{j}\right\rangle N/2-\omega+\epsilon_{j}}\right\rangle }\, ,\label{eq:alpha_s_general}\end{equation}
where $\left\langle \dots\right\rangle =(\sum_{j}\dots)/N$. 
The  sum over $j$ depends on the particular form of the inhomogeneities of $\epsilon_{j}$ and $g_{j}$.
To be specific, 
we consider the following  limiting cases when $\epsilon_{j}$ varies on a much longer or shorter length scale than $g_{j}$, which also includes the case with either $\epsilon_{j}$ or  $g_{j}$ being constant.
In this case and for  large $N$ the sum can be substituted by an integral,
$(\sum_{j}\dots)/N\rightarrow\int d\epsilon dg\ P\left(\epsilon\right)Q\left(g\right)$,
where $P\left(\epsilon\right)$ and $Q\left(g\right)$ are independent normalized distribution functions of the
excitation energies and coupling constants, respectively.
The integral  over $g$ in  Eq. (\ref{eq:alpha_s_general})  separates
and gives an
effective coupling $\sqrt{\left\langle g^{2}\right\rangle}$ \cite{distribution_functions}. 
Further, 
we assume that the boson mode $\omega$ is tuned in
resonance with the spin bath, i.e. $\omega-\left\langle \epsilon\right\rangle =0$. 

\section{Inverse Laplace transform\label{III}}
The inverse Laplace transform of Eq. (\ref{eq:alpha_s_general}) depends
on the particular form of $P\left(\epsilon\right)$ that determines the
analytic structure of $\alpha\left(s\right)$. We will analyze several
cases below. 
If the spin bath is homogeneous then $P\left(\epsilon\right)=\delta\left(\epsilon-\omega\right)$, and $\alpha\left(s\right)$ has two poles on the imaginary axis at
$s=\pm i\sqrt{N\left\langle g^{2}\right\rangle }$, with the associated residues $1/2$. 
In the time domain these poles give 
$\alpha\left(t\right)=\cos(\Omega t)$, where $\Omega=\sqrt{N\left\langle g^{2}\right\rangle }$
is the collective Rabi frequency due to all $N$ spins. This agrees with the result obtained from
exact diagonalization  \cite{CummingsDorri}. 

Next, we consider an inhomogeneous spin bath with excitation energies
spread over a band of width $\Delta$, for which we have $P\left(\epsilon\right)
=\theta\left(-\epsilon+\omega+\Delta/2\right)
\theta\left(\epsilon-\omega+\Delta/2\right)/\Delta$,
where $\theta\left(x\right)$  is the step function. This case is  realized e.g. for 
$\epsilon_{j}=j\Delta/N$, $-N/2\leq j\leq N/2$, i.e. spins in a magnetic field with constant gradient.
 The integral over  $\epsilon$ 
in Eq. (\ref{eq:alpha_s_general}) gives
\begin{equation}
\alpha\left(s\right)=\frac{i}{is+\frac{N\left\langle g^{2}
\right\rangle }{\Delta}\ln\left(\frac{i s-\Delta/2}{i s+\Delta/2}\right)}.\label{eq:alpha_s_no_detuning}\end{equation}
Note that the inverse Laplace transform of Eq. (\ref{eq:alpha_s_general})
is in principle a quasi-periodic function of $t$. Therefore, Eq. (\ref{eq:alpha_s_no_detuning})
is correct  up to the Poincare recurrence time $t_{p}$ which we can estimate as follows.
 We evaluate
the discrete sum over $\epsilon_{j}$ exactly, expand it in $1/N$, and
estimate the time at which corrections to the logarithmic term in
Eq. (\ref{eq:alpha_s_no_detuning}) (due to discretness of the sum)
become important to be $t_{p}=N/\Delta$. Thus, the following time behavior
is valid for times less than $t_{p}=N/\Delta$. For small $N$ it is more convenient to find the few poles of Eq. (\ref{eq:alpha_s_general}) directly and analyze  $\alpha(t)$ numerically as a sum of few harmonic modes rather than to use Eq. (\ref{eq:alpha_s_no_detuning}).

We discuss now the analytic structure of $\alpha\left(s\right)$ in Eq. (\ref{eq:alpha_s_no_detuning}).
There are two branch points at $s=\pm i\Delta/2$ due to the logarithm.
We choose the branch cut as a straight line between these two points.
In addition, there are two poles at $s=\pm is_{0}$ given by the zeroes of the denominator
where $s_{0}$ is a real and positive solution of \begin{equation}
\exp\left(-\frac{s_{0}\Delta}{N\left\langle g^{2}\right\rangle }\right)=\frac{s_{0}-\Delta/2}{s_{0}+\Delta/2}.\label{eq:eq_s0}\end{equation}
In the time domain, the amplitude  $\alpha$
has two contributions, $\alpha=\alpha_{p}+\alpha_{c}$.
One is given by the poles, \begin{equation}
\alpha_{p}\left(t\right)=\frac{2}{1+N\left\langle g^{2}\right\rangle /\left(s_{0}^{2}-\Delta^{2}/4\right)}\cos\left(s_{0}t\right).\label{eq:alpha_p}\end{equation}
This contribution describes a residual oscillation at long times
with amplitude that is reduced from the initial value $\alpha\left(0\right)=1$.
The other one is given by the integral enclosing the branch cut, 
\begin{equation}
\alpha_{c}\left(t\right)=
\int_{0}^{1}dy  \frac{{(4N\left\langle g^{2}\right\rangle }/{\Delta^{2}})\,\cos\left(y\Delta t/2\right)}{\left(y-\frac{2N\left\langle g^{2}\right\rangle }{\Delta^{2}}\ln\left(\frac{1+y}{1-y}\right)\right)^{2}+\left(\frac{2\pi N\left\langle g^{2}\right\rangle }{\Delta^{2}}\right)^{2}} .\label{eq:alpha_c_general}\end{equation}
This contribution describes the decay that occurs due to destructive
interference of many modes forming a continuous spectrum (for large N).

\begin{figure}[pt]
\centering\includegraphics[width=1\columnwidth]{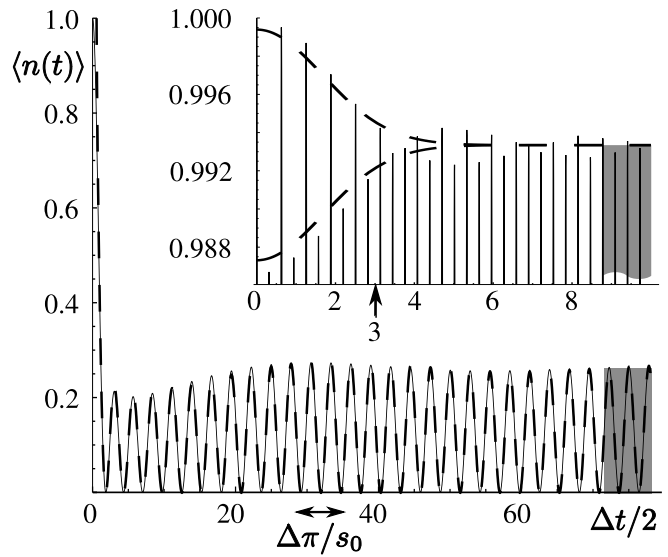}
\caption{Time evolution of the boson $\langle n(t) \rangle=|\alpha(t)|^2$ obtained from numerical evaluation of 
Eqs.~(\ref{eq:alpha_p}, \ref{eq:alpha_c_general}) - full lines. Period of oscillation is $T=2\pi/s_0$  and grey bars are $|\alpha_p(0)|^2$.
Main plot: $\Delta/\Omega=2.2$; dashed line: $|\alpha_p(t)+\alpha_c(t)|^2$ from Eqs.~(\ref{eq:alpha_p}, \ref{eq:alpha_s_CiSi}), $s_0=0.57\Delta$, and  $|\alpha_p(0)|^2=0.26$. Inset: $\Delta/\Omega=0.2$; dashed line: $|\alpha_p(0)\pm\alpha_c(t)|^2$ from Eqs.~(\ref{eq:alpha_p}, \ref{eq:alpha_c_gaussian}) (valid for $\Delta t/2<3$) and from Eq.~(\ref{eq:alpha_s_CiSi}) (valid for $\Delta t/2>3$). The decay of $|\alpha(t)|^2$ is small, ($|\alpha_p(0)|^2=1-\Delta^2/6\Omega^2$), 
and the main contribution  comes from $\alpha_p$. 
}
\label{fig1}
\end{figure}

The integral in Eq. (\ref{eq:alpha_c_general}) can be  approximated
quite accurately for $t\gg2/\Delta$. Due to the fast oscillating cosine the main contribution to the integral
comes from $y\lesssim 2/\Delta t\ll1$. 
Expansion of the logarithm in 
Eq. (\ref{eq:alpha_c_general})
for small $y$ permits us to evaluate the integral in terms of the Integral
Sine and Cosine. An expansion of these special functions for 
$\Delta t/2\gg1$ gives \begin{equation}
\alpha_{c}\left(t\right)=\frac{\Delta^{2}}{N\left\langle g^{2}\right\rangle }\left(\frac{Ae^{-A\Delta t/2}}{2\pi}+\frac{A^{2}\sin\left(\Delta t/2\right)}{\pi^{2}\left(1+A^{2}\right)\Delta t/2}\right)\label{eq:alpha_s_CiSi}\, , \end{equation}
where $A=\pi/2/\left|1-\Delta^{2}/4N\left\langle g^{2}\right\rangle \right|$. Note that for vanishing coupling $g$, 
$\alpha_{p}\left(t\right)$ vanishes and $\alpha_{c}\left(t\right)$ tends to one.
Further, the integrand in Eq.  (\ref{eq:alpha_c_general}) can be expanded for $\Delta^2/N\left\langle g^{2}\right\rangle$ for $\Delta^2\ll N\left\langle g^{2}\right\rangle$. The leading term is linear in $\Delta^2/ N\left\langle g^{2}\right\rangle$ and the remaing integral in the prefactor is a complicated decaying function of $t$ which we approximate qualitatively. First, we perform a change of variable 
- $y=\tanh(x)$ 
turning the  denominator into  
$1/f=\exp\left(-\log\left(f\right)\right)$,
where we expand $\log\left(f\right)$ up to $x^{2}$
and linearize the argument of the cosine in $x$ for $x\ll1$. Finally,
as a result of the Gaussian integral over $x$ we obtain a Gaussian
decay law,
\begin{equation}
\alpha_{c}\left(t\right)=\frac{\Delta^{2}}{2\pi N\left\langle g^{2}\right\rangle }\sqrt{\frac{\pi}{\pi^2+4}}\exp\left(-\frac{\pi^2\Delta^{2}t^{2}}{4(\pi^2+4)}\right).
\label{eq:alpha_c_gaussian}\end{equation}
This approximation agrees reasonably well with  Eq.  (\ref{eq:alpha_c_general}) when evaluated numerically for $t<6/\Delta$ but  breaks down for $t>6/\Delta$ where Eq.  (\ref{eq:alpha_s_CiSi}) is valid, see Fig. \ref{fig1}.

The time-dynamics of 
$\left\langle n\left(t\right)\right\rangle =|\alpha\left(t\right)|^{2}$
can be classified in terms of the ratio $\Omega/\Delta$, 
with Rabi frequency $\Omega=\sqrt{N\left\langle g^{2}\right\rangle }$.
If the inhomogeneity of the spin bath is small, $\Delta\ll\Omega$, the boson
oscillates with a single frequency like in the  homogeneous
case. The main contribution to $\alpha\left(t\right)$ comes from
the poles Eq. (\ref{eq:alpha_p}) with $s_{0}=\Omega+\Delta^2/24\Omega$, 
which is shifted  with respect to the homogeneous system. The amplitude
of $\alpha\left(t\right)$ is only slightly reduced from its initial
value, $1-\Delta^{2}/12\Omega^{2}$. The decay law to this value is mainly Gaussian-like,
Eq. (\ref{eq:alpha_c_gaussian}), with the decay time $t_{1}\approx2.4/\Delta$, see Fig. \ref{fig1}.
If the spin bath is strongly inhomogeneous, $\Delta\gg\Omega$, the 
boson mode decays completely from $\alpha\left(0\right)=1$
to $0$. The main contribution to $\alpha\left(t\right)$ comes from
the branch cut, Eq. (\ref{eq:alpha_s_CiSi}), with $A\approx2\pi\Omega^{2}/\Delta^{2}$,
whereas the pole contribution is exponentially small. The decay 
behavior is mainly exponential with timescale $t_{2}\approx\Delta/\pi\Omega^{2}$.
At long times $t\gg t_{2}$ the second term in Eq. (\ref{eq:alpha_s_CiSi})
becomes dominant, exhibiting a slow power-law decay. 

In the intermediate regime, $\Delta\simeq\Omega$, the time decay is only partial, with the amplitude
of the residual oscillation of $\alpha\left(t\right)$ being less than unity but staying constant in time.
Its precise value can be found from the numerical solution of Eqs. (\ref{eq:eq_s0},\ref{eq:alpha_p}).
The decay displayed in Eq. (\ref{eq:alpha_s_CiSi}) is governed by a combination of exponential and power law behavior.
As $A\simeq1$ and $s_{0}\simeq\Omega\simeq\Delta$ there is no clear
separation of time scales coming from the exponential, the inverse power law and the oscillatory contribution, see Fig. \ref{fig1}.
Note that in case of $\Delta=2\Omega$ the first term
in Eq. (\ref{eq:alpha_s_CiSi}) vanishes, thus the decay in this particular case is purely
power law. The non-standard dynamics, in particular the  non-exponetial decay in the intermediate regime, 
is a manifestation of the quantum nature of the system. For other models with non-Markovian decay see e.g.  \cite{Leggett,Khaetskii}.

For a Gaussian distribution $P\left(\epsilon\right)=\exp(-\left(\epsilon-\omega\right)^{2}/\Delta^{2})/\sqrt{\pi}\Delta$ 
the dynamics we find is
qualitatively the same as the one obtained before 
for the
rectangular  distribution, see Appendix A. 
The $\epsilon$-integral 
 in Eq. (\ref{eq:alpha_s_general}) 
 leads to
 the complex error function of $s$.
In Laplace
space, $\alpha\left(s\right)$ exhibits one branch cut along the imaginary
axis that vanishes at  $\pm i\infty$. In the time
domain, $\alpha\left(t\right)$ is given by an integral around this
branch cut. In the limit of $\Delta\ll\Omega$ we recover the previous
result for the homogeneous spin bath. In the opposite limit of 
strong inhomogeneity, $\Delta\gg\Omega$, we obtain the same result
as in Eq. (\ref{eq:alpha_s_CiSi}) 
up to numerical prefactors $\sqrt{\pi}$. 

The physical interpretation of the decay is as follows.
The boson  flips, say, spin $j$, and then this spin precesses
for some time with a frequency $\epsilon_{j}$ before
this excitation gets transfered back to the boson. The acquired phase of the boson is thus different for each particular spin. The sum over these random phases eventually leads to  destructive interference (for $N\gg 1$) and thus to a decay.

\section{Finite detuning\label{IV}}
Next, we analyze the effect of finite detuning. If the spin bath is homogeneous,
a small detuning $\left|\omega-\epsilon\right|\ll\Omega$ forces $\alpha\left(t\right)$
to oscillate with two distinct frequencies $\left(N-1\right)\left(\omega-\epsilon\right)/2\pm\sqrt{\left(\epsilon-\omega\right)^{2}+\Omega}/2$
instead of only one $\Omega$. A large detuning $\left|\omega-\epsilon\right|\gg\Omega$
suppreses the dynamics of $\alpha\left(t\right)$. The phase of the wave
function oscillates with frequency $N\left(\omega-\epsilon\right)/2$
but the amplitude stays constant at the initial value of $\alpha\left(0\right)=1$
up to a small correction of the order of $\Omega^{2}/\left(\omega-\epsilon\right)^{2}$.

In the inhomogeneous case we perform a similar calculation
as for zero-detuning and obtain  $\alpha\left(s\right)$
with an analytic structure similar to Eq. (\ref{eq:alpha_s_no_detuning}), see Appendix B.
There are two poles  on the imaginary axis and a branch cut
that is responsible for the relaxation. Explicit expressions for $\alpha_{p}\left(t\right)$
and $\alpha_{c}\left(t\right)$ can be obtained and are generalizations of Eqs. (\ref{eq:alpha_p},\ref{eq:alpha_c_general}), see Eq. (\ref{alpha_s_detuning}). For small detuning $\left|\left\langle \epsilon\right\rangle -\omega\right|\ll\Omega$
two poles emerge that  are not complex conjugates of each other and thus lead to
two distinct frequencies of the residual oscillations of $\alpha\left(t\right)$.
Large detuning $\left|\left\langle \epsilon\right\rangle -\omega\right|\gg\textrm{max}\{\Delta,\Omega\}$
suppresses  the relaxation and any long-time dynamics. The main contribution to $\alpha\left(t\right)$
comes from one of the poles with residue $1-\Omega^{2}/\left(\omega-\left\langle \epsilon\right\rangle \right)^{2}$.
Thus, the initial value $\alpha\left(0\right)=1$ remains
almost unaltered under evolution independent of the ratio $\Omega/\Delta$.

The dynamics at large detuning can also be analyzed using perturbation
theory. Applying a Schrieffer-Wolff transformation to the Dicke Hamiltonian
the  boson-spin coupling can be removed to lowest order in $g$
and thereby an effective XY spin-coupling within the spin bath is obtained \cite{Trifa:2008}.
As a result, 
the boson  number $n$ and the z-component
of the total spin $\sum_{j}S_{j}^{z}$
are conserved separately by this effective Hamiltonian. Thus, again, the initially
excited boson mode will remain unaltered under the evolution in
leading order of the perturbation. However, there is a
virtual boson process which induces the dynamics within the spin bath.

\section{Conclusions}
In conclusion, we analyzed the dynamics of a single boson
mode coupled to an inhomogeneous spin bath exactly, and found
a complex decay behavior of the boson.
While we focused in this work on particular inhomogeneities of the spin bath excitation energies, 
it is straightforward to apply the approach presented here  to other cases.

\section{Aknowledgments}
We thank  M. Duckheim and M. Trif for discussions. We acknowledge
support from the Swiss NSF, NCCR Nanoscience Basel, JST
ICORP, and DARPA Quest.

\appendix
\section{Gaussian distribution of the spins' splitting energies}
Here we derive the time dynamics resulting from the Gaussian distribution function of $\epsilon$, $P\left(\epsilon\right)=\frac{1}{\sqrt{\pi}\Delta}e^{-\left(\epsilon-\omega\right)^{2}/\Delta^{2}}$.
Performing the integral over $\epsilon$ in Eq. (\ref{eq:alpha_s_general}) we obtain \begin{equation}
\alpha(s)=\frac{1}{s+\frac{\sqrt{\pi}N\left\langle g^{2}\right\rangle }{\Delta}\varpi\left(\imath\frac{s}{\Delta}\right)},\label{eq:alpha_s_gauss}\end{equation}
where $\varpi\left(z\right)$ is defined in the upper and lower complex half-planes
separately as \begin{equation}
\varpi\left(z\right)=\begin{cases}
w\left(z\right) & ,\textrm{Im}\; z\ge0\\
-w\left(-z\right) & ,\textrm{Im}\; z<0\end{cases}\end{equation}
and where  $w\left(z\right)=e^{-z^{2}}\textrm{erfc}\left(-\imath z\right)$ is the error function.
The function $\varpi\left(z\right)$ has a branch cut along the real axis,
$\lim_{\delta\rightarrow0}\varpi\left(\pm\imath\delta\right)=\pm1$, which 
vanishes at infinity,
\begin{equation}
\lim_{x\rightarrow\pm\infty}\lim_{\delta\rightarrow0}\varpi\left(x\pm\imath\delta\right)=\lim_{x\rightarrow\pm\infty}e^{-x^{2}}\left(\pm1+\textrm{erf}\left(\imath x\right)\right)=0.\end{equation}

The inverse Laplace transform is given by an integral around the entire imaginary
axis 
\begin{widetext}\begin{eqnarray}
\alpha\left(t\right) & = & -\frac{N\left\langle g^{2}\right\rangle }{2\sqrt{\pi}\Delta^{2}}\int_{-\infty}^{\infty}dy\frac{e^{-\imath yt/2\Delta}e^{-4y^{2}}}{\left(\imath y-\frac{\sqrt{\pi}N\left\langle g^{2}\right\rangle }{2\Delta^{2}}w\left(2y\right)\right)^{2}+\frac{\sqrt{\pi}N\left\langle g^{2}\right\rangle e^{-4y^{2}}}{\Delta^{2}}\left(\imath y-\frac{\sqrt{\pi}N\left\langle g^{2}\right\rangle }{2\Delta^{2}}w\left(2y\right)\right)}\label{eq:alpha_case_c}\end{eqnarray}
\end{widetext} where the substitutions $s=\imath2\Delta y$ and $\omega\left(-z\right)=2e^{-z^{2}}-\omega\left(z\right)$
were used.

For $\Delta\ll\sqrt{N\left\langle g^{2}\right\rangle }$  Eq. (\ref{eq:alpha_s_gauss}) can be expanded in the small parameter $\Delta $. The leading term has an analytical structure similar
to Eq. (\ref{eq:alpha_s_no_detuning}). There are two symmetric poles on the imaginary axis and
a finite length branch cut between $s=\pm\imath2\Delta$. The contribution from the poles is \begin{equation}
\alpha_{p}\left(t\right)=\frac{2\cos\left(s_{0}t\right)}{1-N\left\langle g^{2}\right\rangle /s_{0}^{2}}.\end{equation}
Using the large $z$ asymptotics of the error function $\textrm{erfc}\left(z\right)=\frac{e^{-z^{2}}}{\sqrt{\pi}z}\left(1-\frac{1}{2z^{2}}\right)$,
the two poles are given by $s_{0}=\pm\imath\sqrt{N\left\langle g^{2}\right\rangle }$.
The residues at this poles are \begin{equation}
\underset{s=s_{0}}{\textrm{Res}}\;\alpha(s)e^{st}=\frac{e^{s_{0}t}}{2}.\end{equation}
Thus, the contribution from the poles is dominant. In this limit we recover
the non-interacting case, a single Rabi oscillation,\begin{equation}
\alpha\left(t\right)=\cos\left(\sqrt{N\left\langle g^{2}\right\rangle }t\right).\end{equation}

In the opposite regime $\Delta\gg\sqrt{N\left\langle g^{2}\right\rangle }$ 
 there are no poles  and there is just a single branch cut. The long
time asymptotics can be evaluated by expanding the denominator for small
$y$ and approximating $e^{-4y^{2}}\approx 1$ in the numerator,
\begin{eqnarray}
\alpha_{c}\left(t\right) & = & \frac{N\left\langle g^{2}\right\rangle }{\sqrt{\pi}\Delta^{2}}\int_{0}^{1}dy\frac{\cos\left(yt/2\gamma\right)}{y^{2}+\left(\frac{\sqrt{\pi}N\left\langle g^{2}\right\rangle }{2\Delta^{2}}\right)^{2}}.
\end{eqnarray}
This integral, up to a numerical factor, is the same  as in Eq. (\ref{eq:alpha_c_general}) in this limit. 

\section{Calculation for $\langle\epsilon\rangle\neq\omega$}
Here we assume that the detuning is finite $\gamma=\left\langle \epsilon\right\rangle -\omega\neq0$.
We repeat the same steps as before and similarly to the zero detuning
case we obtain in the Laplace domain
\begin{equation}
\alpha\left(s\right)=\frac{\imath}{\imath s+\frac{N\gamma}{2}+\frac{N\left\langle g^{2}\right\rangle }{\Delta}\ln\left(\frac{\imath s+\left(N-2\right)\gamma/2-\Delta/2}{\imath s+\left(N-2\right)\gamma/2+\Delta/2}\right)}.\label{alpha_s_detuning}\end{equation}
This function is characterized by
two poles and one branch cut. 

The two poles are given by zeroes of the denominator $s=\imath\left(N\gamma/2+s_{1,2}\right)$
where $s_{1,2}$ are the solutions of 
\begin{equation}
\exp\left(-\frac{s\Delta}{N\left\langle g^{2}\right\rangle }\right)=\frac{s-\gamma-\Delta/2}{s-\gamma+\Delta/2}.\label{eq:s_01_h}
\end{equation}
This equation is not symmetric with respect to  $s\rightarrow-s$,
thus the two poles are not symmetric. The residues of the poles are given by
\begin{equation}
\underset{s}{\textrm{Res}}\alpha_{s}e^{st}=\frac{1}{1+\frac{N\left\langle g^{2}\right\rangle }{\left(s_{1,2}-\gamma\right)^{2}-\Delta^{2}/4}}e^{st}.\end{equation}
Performing the inverse Laplace transformation we obtain similarly
to Eq. (\ref{eq:eq_s0}) \begin{equation}
\alpha_{p}\left(t\right)=\sum_{k=1,2}\frac{e^{\imath N\gamma t/2+\imath s_{k}t}}{1+\frac{N\left\langle g^{2}\right\rangle }{\left(s_{k}-\gamma\right)^{2}-\Delta^{2}/4}}.\label{eq:alpha_p2}\end{equation}
The  branch points are $s=\imath\left(N\gamma/2\pm\Delta/2\right)$.
Similarly to the case of zero detuning the contribution from the branch
cut is given by the integral \begin{widetext}\begin{eqnarray}
\alpha_{c}\left(t\right) & = & \frac{2N\left\langle g^{2}\right\rangle e^{\imath\left(\left(N-2\right)\gamma+\Delta\right)t/2}}{\Delta^{2}}\int_{-1}^{1}dy\frac{e^{\imath y\Delta t/2}}{\left(y-\frac{2\gamma}{\Delta}-\frac{2N\left\langle g^{2}\right\rangle }{\Delta^{2}}\ln\left(\frac{1+y}{1-y}\right)\right)^{2}+\left(\frac{2\pi N\left\langle g^{2}\right\rangle }{\Delta^{2}}\right)^{2}}.\end{eqnarray}
\end{widetext}

At  small detuning $\left|\omega-\left\langle \epsilon\right\rangle \right|\ll\textrm{max}\left(\Delta/2,N\left\langle g^{2}\right\rangle \right)$
there are two distinct frequencies in Eq. (\ref{eq:alpha_p2}), thus
the final state oscillates with two frequencies. For a large detuning
$\left|\omega-\left\langle \epsilon\right\rangle \right|\gg\textrm{max}\left(\Delta/2,N\left\langle g^{2}\right\rangle \right)$
the relaxation is suppressed. In the limit of strong detuning the roots
of Eq. (\ref{eq:s_01_h}) are given by $s_{1}=-2N\left\langle g^{2}\right\rangle /\gamma$
and $s_{2}=\gamma-\Delta/2$. The residue at $s_{2}$ is exponentially
small and the contribution from the poles is given only by the pole $s_{1}$

\begin{equation}
\alpha_{p}\left(t\right) = \frac{e^{-\imath2N\left\langle g^{2}\right\rangle t/\Delta+\imath N\gamma t/2}}{1+\frac{N\left\langle g^{2}\right\rangle }{\left(2N\left\langle g^{2}\right\rangle /\gamma+\gamma\right)^{2}-\Delta^{2}/4}}  \approx  e^{\imath N\gamma t/2}.\end{equation}
In this result the amplitude of $\alpha\left(t\right)$ remains constant in time. From the initial condition $\alpha_{p}\left(0\right)+\alpha_{c}\left(0\right)=1$
the contribution from the branch cut is negligible, and therefore there is no decay
for sufficiently strong detuning. The corrections to this result are small and of the order of
$\textrm{max}\left(\Delta/2,N\left\langle g^{2}\right\rangle \right)/\left|\omega-\left\langle \epsilon\right\rangle \right|$.


\begin{thebibliography}{10}
\bibitem{Gardiner}C. W. Gardiner and P. Zoller, \emph{Quantum Noise},
Springer, 2004.

\bibitem{cold_atoms_BEC}K. B. Davis, M. -O. Mewes, M. R. Andrews,
N. J. van Druten, D. S. Durfee, D. M. Kurn, and W. Ketterle, Phys.
Rev. Lett. \textbf{75}, 3969 (1995)

\bibitem{kimble}T. Aoki, B. Dayan, E. Wilcut, W. P. Bowen, A. S.
Parkins, T. J. Kippenberg, K. J. Vahala, and H. J. Kimble, Nature
\textbf{443}, 671 (2006).

\bibitem{esslinger}F. Brennecke, T. Donner, S. Ritter, T. Bourdel,
M. K\"{o}hl, and T. Esslinger, Nature \textbf{450}, 268 (2007).

\bibitem{Rempe}G. Rempe, H. Walther, and N. Klein, Phys. Rev. Lett.
\textbf{58}, 353 (1987).

\bibitem{Bloch}M. Greiner, O. Mandel, T. W. H\"{a}nsch, and I. Bloch,
Nature 419, \textbf{51} (2002).

\bibitem{Wineland}
D. J. Wineland, J. J. Bolinger, W. M. Itano, and D. Heinzen,
Phys. Rev. A \textbf{50}, 67 (1994).
 
\bibitem{Kasevich}
A. K. Tuchman, R. Long, G. Vrijsen, J. Boudet, J. Lee, and M. A. Kasevich,
Phys. Rev. A \textbf{74}, 053821 (2006).

\bibitem{kasprzak}J. Kasprzak, M. Richard, S. Kundermann, A. Baas,
P. Jeambrun, J. M. J. Keeling, F. M. Marchetti, M. H. Szyma\'{n}ska,
R. Andr\'{e}, J. L. Staehli, V. Savona, P. B. Littlewood, B. Deveaud,
and Le Si Dang, Nature \textbf{443}, 409 (2006).

\bibitem{ep3}R. Balili, V. Hartwell, D. Snoke, L. Pfeiffer, and K. West,
Science \textbf{316}, 1007 (2007).

\bibitem{wallraff}A. Wallraff, D. I. Schuster, A. Blais, L. Frunzio,
R.- S. Huang, J. Majer, S. Kumar, S. M. Girvin, and R. J. Schoelkopf,
Nature \textbf{431}, 162 (2004) .

\bibitem{Berezovsky}
J. Berezovsky, M. H. Mikkelsen, N. G. Stoltz, L. A. Coldren, and D. D. Awschalom, Science \textbf{320}, 349 (2008). 

\bibitem{Badolato1}K. Hennessy, A. Badolato, M. Winger, D. Gerace,
M. Atat\"{u}re, S. Gulde, S. F\"{a}lt, E. L. Hu, and A. \.{I}mamo\u{g}lu,
Nature \textbf{445}, 896 (2007).

\bibitem{Cirac}
J. I. Cirac and P. Zoller, Phys. Rev. Lett. \textbf{74}, 4091 (1995).

\bibitem{AwschalomQED} A. \.{I}mamo\u{g}lu, D. D. Awschalom, G.
Burkard, D. P. DiVincenzo, D. Loss, M. Sherwin, and A. Small, Phys.
Rev. Lett. \textbf{83}, 4204 (1999).


\bibitem{Lukin}L. Childress, A. S. Sorensen, and M. D. Lukin, Phys. Rev. A \textbf{69}, 042302 (2004).
 
\bibitem{Burkard} G. Burkard and A. Imamoglu, Phys. Rev. B \textbf{74}, 041307R (2006).

\bibitem{Trifa:2008}
M. Trif, V. N. Golovach, and D. Loss, Phys. Rev. B
\textbf{77}, 045434 (2008).

\bibitem{Dicke}R. H. Dicke, Phys. Rev. \textbf{93}, 99 (1954).

\bibitem{TavisCummings}M. Tavis and F. W. Cummings, Phys. Rev. 170,
379 (1968).

\bibitem{CummingsDorri}F. W. Cummings and A. Dorri, Phys. Rev. A
\textbf{28}, 2282 (1983).

\bibitem{Chumakov}M. Kozierowski, A. A. Mamedov, and S. M. Chumakov,
Phys. Rev. A \textbf{42}, 1762 (1990); I. Sainz, A. B. Klimov, and
S. M. Chumakov, J. Opt. B: Quantum Semiclass. Opt. \textbf{5}, 190(2003).

\bibitem{Solano} C. E. Lopez, H. Christ, J. C. Retamal, and E. Solano, Phys. Rev. A \textbf{75}, 033818 (2007).

\bibitem{ExcitonPolaritonsEstimate} For an exciton-polariton system the spin is electron-hole excitation. In \cite{kasprzak} $\Omega\simeq 26$ meV and $\omega\simeq 1.7$ eV. Disorder in a semiconductor can be 0.1-50 meV.

\bibitem{ColdGasesEstimate} For a cold gas system single spin is a hyperfine state of an atom. In \cite{esslinger} $\Omega\simeq 2$ GHz and $\omega\simeq 7$ GHz. In  \cite{Kasevich} $\Omega\simeq 10$ MHz and $\omega\simeq 0.7$ GHz. From  $10$ T gradient of the magnetic field  inhomogeneity can be estimated $\Delta\simeq 10$ MHz.

\bibitem{YamamotoImamoglu} Y. Yamamoto and A. Imamoglu, \textit{Mesoscopic Quantum Optics}, John Willey and Sons, Inc., 1999.

\bibitem{applicability}The Dicke model is valid near the resonance
between boson and spin bath  energies, $\left|\omega-\epsilon_{j}\right|\ll\omega,\epsilon_{j}$.
Still, if $\Delta\ll\omega$, a relatively large detuning $\left|\left\langle \epsilon\right\rangle -\omega\right|>\Delta,\Omega$ can also be studied within this model.

\bibitem{superradiance} For high temperatures,
 $T\gg N\epsilon$, thermal spin excitations get transfered collectively to the boson. Such a spontaneous high population ($n\gg 1$) of a photon mode leads to the Dicke `superradiance' effect \cite{Dicke}.
 
 \bibitem{centralspin} A similar ansatz is used in the central spin model describing
the  inhomogeneous isotropic interaction between a single electron spin and a nuclear spin bath \cite{Khaetskii}.

\bibitem{Khaetskii}
A. Khaetskii, D. Loss, and L. Glazman,
Phys. Rev. Lett. \textbf {88}, 186802 (2002).


\bibitem{distribution_functions} If $g_{i}$ and $\epsilon_{i}$ are correlated, e.g. $P(\epsilon,g)=\delta(\epsilon-g) \theta(-\epsilon+\omega+\Delta/2) \theta(\epsilon-\omega+\Delta/2)/\Delta$, the integrals over $g$ and $\epsilon$ do not separate in Eq. (\ref{eq:alpha_s_general}) even for
$N\gg 1$. 
 
\bibitem{Leggett} A. J. Leggett, S. Chakravarty, A. T. Dorsey, M. P. A. Fisher, A. Garg, and W. Zwerger, Rev. Mod. Phys. \textbf{59}, 1 (1987). 

 
\end{thebibliography}
\end{document}